# Birth of Universe following Rupture of Fractal Structure Filaments

**V.A. Rantsev-Kartinov**

*Kurchatov Institute,
Kurchatov Square 1,
123182, Moscow, Russia*
e-mail: rank@nfi.kiae.ru

**C.G. Parigger***

*The University of Tennessee Space Institute,
411 B.H. Goethert Parkway,
Tullahoma, TN 37388, U.S.A.*
e-mail: cparigge@tennessee.edu
*Corresponding author.

*This work addresses unique characteristics of our Universe, namely fractal structures that consist of coaxial-tubular and wheel-like building blocks. The front view of the tubes may show a wheel, and in turn, the wheel may show radial tubes arranged as spokes. A flexible interconnect originates from magnetic fields. Neutrino astronomy indicates fractal structure in the core of stars and galaxies. Fractals are fundamental for modeling baryonic filaments important in "dark matter" with a topology that can be inferred from radiating cosmic objects. The rupture of dark-matter filaments may cause formation of cosmic objects including birth of the Universe.*

# I.  INTRODUCTION

Filament structures were discovered in laboratory thermo-nuclear plasmas that persisted over time scales larger than predicted by magnetic hydrodynamics and physical kinetics computations [1, 2]. These filament structures consist of separate identical blocks linked together in a network of: (i) Coaxial-tubular structures with internal radial links, and (ii) wheel-like structures which exist either separately or occur at the end of the tubes. Figure 1 illustrates the inter-connected manifold. The cosmic wheel's skeleton is repeated as indicated.

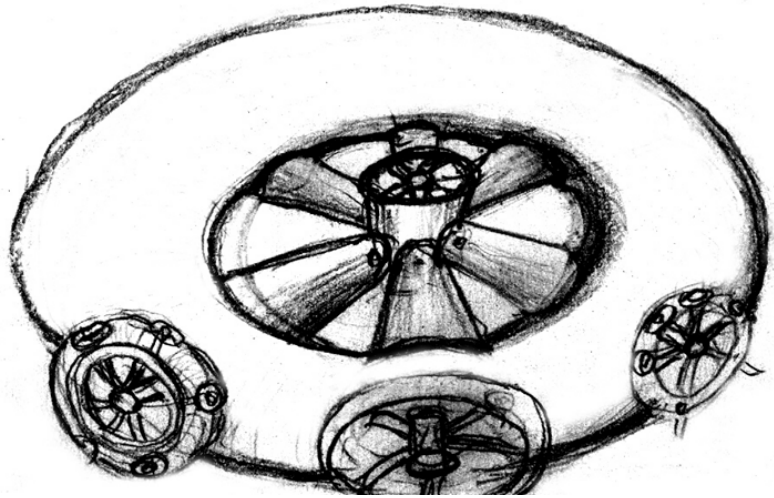

**Figure 1.** Schematic "tube" and "wheel" structure.

Plasma images indicate self-similarity of the observed structure, i.e., details of the whole show fractal characteristics. A multilevel dynamical contrast (MDC) method [2-4] allows us to investigate embedded structures inherent in data, even at low signal-to-noise levels. Repeated plasma observations and use of MDC image enhancement lead to the following conclusion: Only a micro/nano dusty component of the discovered filaments with its quantum connections can provide an explanation for the persistent structure. Carbon nano-tubes (CNT) were considered as the elementary building blocks [4, 5]. Carbon in form of CNT in plasma naturally occurs due to operation of vacuum pumps, out-gassing processes of plasma device walls and initiation of inductive discharges. Filaments may occur including shielding of the micro-solid skeleton from the plasma. We named the sustained plasma structures "wild cables" initially [6]. The persistence of the micro-solid-state plasma structure has been investigated further in dust sediments on chamber surfaces of plasma devices [7-9]. The deposits primarily consist of amorphous carbon and hydrocarbons; however, the skeleton-like self-similar structures of CNT have been observed in almost all studied sediments collected from the chamber walls. Generations of CNTs were observed at spatial scales as small as several nanometers. In a completed effort regarding these filaments, we reported fractal characteristics of these structures from 0.01 to 0.1-m plasma.

We focused on images of atmospheric phenomena in dusty/aerosol plasma: Tornados, waterspouts and tropical hurricanes [10-12] in our study of larger fractal size and/or topology. In addition, ocean skeleton-structures of the same topology have been investigated [13]. These research efforts have allowed us to explore fractals of observable structures at scales in the order of $10^6$ m. Clearly, persistent fractal structures of the same topology were observed in various physical phenomena and in various environments on Earth. The co-axial tubular and the wheel structures can be regarded as the building blocks over a spatial range that covers 15 orders of magnitude, i.e., from $10^{-9}$ m in nano-dust to $10^6$ m in atmospheric dusty plasma and the ocean.

Cosmic spatial scales are of interest in investigation of fractal structures [14-17]. An understanding of the fractal composition of the Universe will allows us to also address: (i) Why the visible mass in space amounts to only a few percents of the mass in the Universe; (ii) Why galaxies may be formed in a relatively short time compared to the age of the Universe; (iii) Why cosmic objects such as planetary nebulas or galaxies show a typically narrow size distribution.

## II.     OBSERVATION OF SUN'S FRACTAL STRUCTURE

Images from the Sun recorded at different wavelengths were analyzed initially [18, 19]. For exploration of fractal structures that may occur in the sun, MDC image enhancement was applied. We discovered the following: i) Skeletal structures exist on the Sun and its vicinity that are topology identical to the ones observed on Earth over a wide range of scales; ii) The majority of coronal mass ejections and protuberances of the Sun are induced by tubular filaments. These almost invisible tubular filaments interact with the Sun and its atmosphere. Coronal ejections and protuberances are not caused by internal thermonuclear explosions; iii) A three-dimensional Sun spot structure develops as a result of extruding dark filaments of the Sun during its activity. The observed structure shows a coaxial-tubular skeleton topology identical to the one mentioned previously [19].

Figure 2 illustrates an image of the Sun together with the indicated tubular structure. The width of figure from the Large Angle Spectrometric Coronograph (LASCO) corresponds to $2.5 \times 10^9$ m. An almost invisible coaxial–tubular structure is directed to the left and upward from the Sun. The plasma flow from the Sun occurs as a filament enters the Sun and its atmosphere. The external diameter of the filament is almost 8 Sun diameters of approximately $10^{10}$ m. The image covers 30 Sun diameters, i.e., $4.2 \times 10^{11}$ m. The diameter of the central portion of this coaxial-tubular structure that touches the Sun typically equals the diameter of the Sun. The figure shows ejection with illustrated structure in the interaction region. The figure also shows additional filaments.

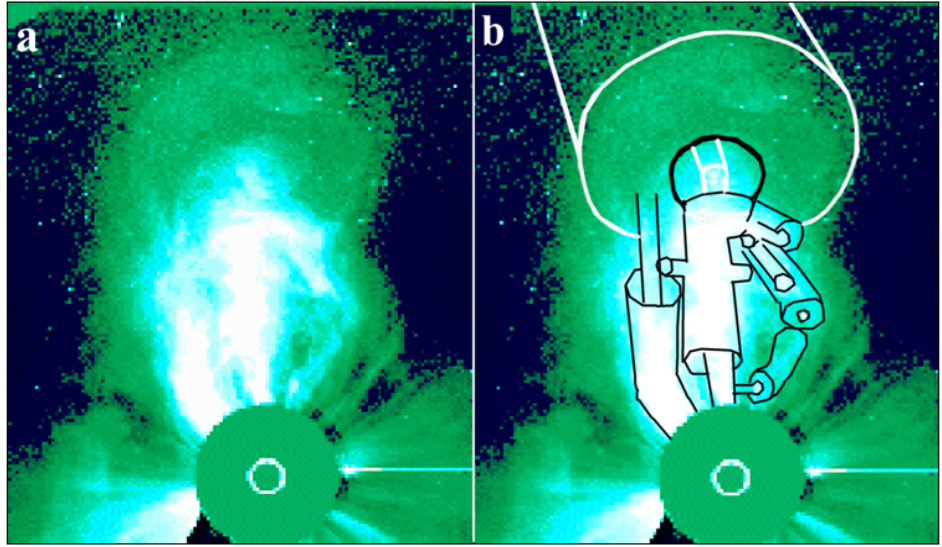

**Figure 2. a.** Image of powerful coronal mass ejections of the Sun analyzed using the multilevel-dynamic contrast method. **b.** Schematic illustration of the tubular structure.

Selected images collected by the Solar Heliospheric Observatory (SOHO) were also analyzed. Figure 3 summarizes the results: i) The left side of Fig. 3 shows a fragment of a Sun's image, recorded in the x-ray range with $\lambda$ = 284 Å (SOHO). The width of image corresponds to $3.6 \times 10^8$ m. The light border at the left elucidates the solar disk. The middle of the image indicates a multi-layer, vertically oriented coaxial-tubular structure (CTS) of a telescopic type with diameter of approximately $1.8 \times 10^8$ m. The bright coaxial tube shows a diameter $2 \times 10^7$ m just inside of the forward part CTS with diameter of $3.6 \times 10^7$ m. The ration of diameters of two neighboring, attached tubes is in the range of 1.6 to 1.8. ii) The center of Fig. 3 shows a section of the image of the Sun in the x-ray range $\lambda$ = 171 Å (SOHO). The width of the figure corresponds to $1.25 \times 10^8$ m. Illustrated are CTS blocks in various positions and joints built from identical but smaller CTS blocks. Notice in the middle of the center images that the CTS angles to the right at 30° from vertical. The CTS diameter is $5 \times 10^7$ m, its length $7 \times 10^7$ m. Diameters of telescopic tubes are indicated at the end of the sketched structures. The ratio of diameters of nearby enclosed tubes amounts to 2. The ends of such tubes highlight the CTS of the building blocks. iii) The right of Fig.3 shows a fragment of the Sun's image in x-ray range $\lambda$ = 195 Å (SOHO). The figure illustrates complex CTS. The edge of a solar disk is illustrated as well. The CTS diameter above the Sun's surface is $2 \times 10^8$ cm. The structure extends to approximately 3 to $4 \times 10^8$ m in height.

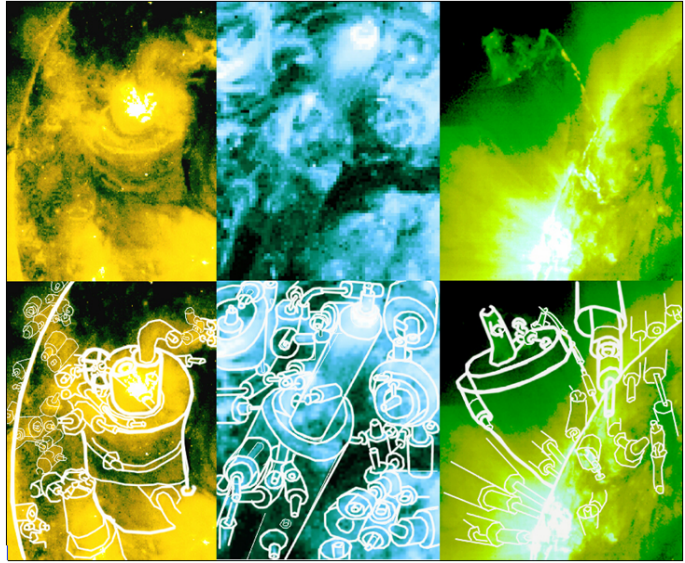

**Figure 3.** Section of the Sun's image collected in the soft-X-ray region. Schematic tubular structures are indicated.

Figure 4 portrays interacting tubular structures. The top of Fig. 4 shows an enhanced image of CTSs of protuberances caused by penetration of an external filament into the Sun's atmosphere and body. The interacting filament stimulates the Sun's activity in this area. The width of the figure amounts to $3.8 \times 10^7$ m. The bottom part of Fig. 4 shows an overlay that highlights the tubular schematics. Diameters of indicated CTSs are of the order $7 \times 10^6$ m.

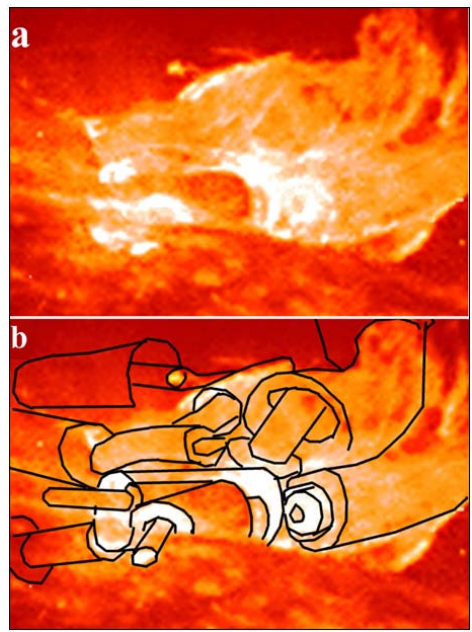

**Figure 4.** Interacting Sun filaments. **a.** MDC-enhanced image. **b.** Enhanced images with illustrations of the interacting tubes.

Selected Sun image sections were further analyzed as well. Figure 5 shows active zones at the Sun's surface together with tube schematics. The skeleton-like structure depicts CTS of the Sun's filaments which extrude from the center of the Sun to its surface. The topology of these structures is the same as observed previously in dust deposit of Tokamak T-10. Displayed are the very details of tri-dimensionality of the structure and its interconnections. Diameters of these CTS amount to $2 \times 10^7$ m. The upper ends extend above the Sun's surface to typically $10^8$ m in height.

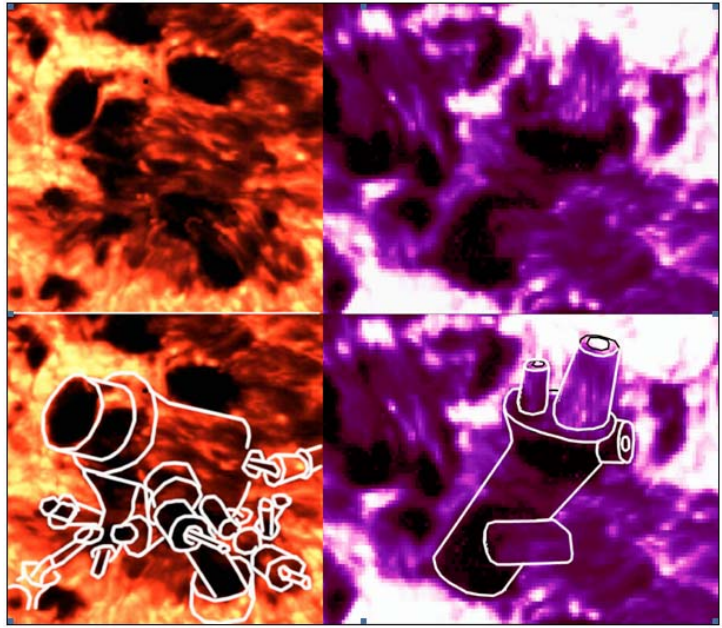

**Figure 5.** Details of image fragments from the Sun's surface together with sketches of the fractal structures.

We summarize the following hypothesis [19, 20]: "The Sun has internal filaments and fractal skeleton structure, which can be a manifestation of one of possible forms of the filaments in matter. The sun-spots are a part of these filaments and skeletal structures, i.e., these are images of expelled filaments that are visible to us when the Sun is active." The fragments of the Sun's internal fractal structure can be an excellent indicator of the phenomena [19] supported by a breadth of observations.

Figure 6 shows a fragment of a Sun's image recorded at $\lambda$ = 195 Å (SOHO) and analyzed with the multilevel dynamical contrast (MDC) method. The edge of a solar disk is recognizable at the top of the image. The complex weaved filament structure shows coaxial-tubular structures (CTS) composed of similar building blocks. Smaller size structures occur along of the Sun's axis. The diameter of CTS amounts to $3.3 \times 10^8$ m, diameters of building blocks of the CTSs are typically $10^7$ m, and the diameter of the filament structure (emanating from the center of its lower end) amounts to $8 \times 10^7$ m. The CTS investigated here shows radius of rotation around of the Sun axis that is slightly smaller than the solar disk radius.

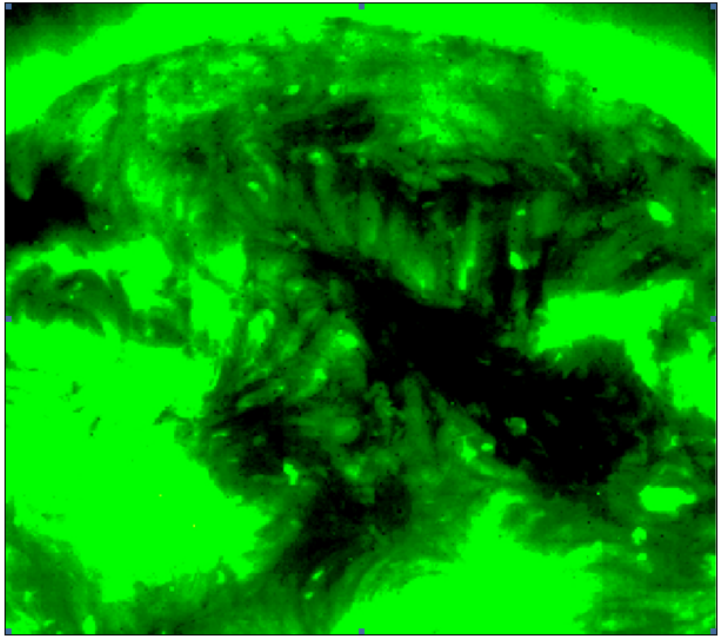

**Figure 6.** Fragment of the Sun's image.

### III.     ANALYSIS OF NEUTRINO SUN IMAGES

Information from within a star can be obtained from neutrinos. Yet it is important to consider what physical processes indeed can be utilized for image formation and reconstruction from within a flux of neutrinos. We first consider a so-called Dirac neutrino, i.e., a neutrino with zero electric charge but non-zero mass. According to Standard model, Dirac neutrinos show non-zero magnetic moment, $\mu_\nu^0 \approx 1.74 \cdot 10^{-23} \cdot m_\nu$ [eV/T], where $m_\nu$ is the neutrino rest-mass [eV].

An image inside a body is generated from absorption and/or dispersion from internal heterogeneities. In the following we investigate processes that can be induced by neutrinos and may be used for neutrino-imaging. It is known, that moving neutrinos show oscillatory behavior even in vacuum [21]. This oscillatory behavior is caused by one type neutrino transforming spontaneously into another type of neutrino. A neutrino in vacuum travels a considerable distance during a characteristic oscillation period. With time, all three neutrino types, namely electronic, muonic and tauonic, will be present in a flux of neutrinos [22]. As neutrinos move in matter with a density gradient, resonant oscillations [21, 22] can be experienced when transitions of one neutrino type into another can take place at only a limited portion of the trajectory. For large density gradients a resonant transition of electronic into muonic neutrino can occur at the characteristic gradient scale. During resonant passage one neutrino type transforms into another, consequently these neutrinos are lost for imaging. Conversely, a portion of the neutrino flux through the internal structures of a star will cause a shadow.

With an average density of the Sun of $10^3$ kg/m$^3$, the electronic number density of plasma amounts to approximately $n_e^S \sim 5 \times 10^{30}$ m$^{-3}$, and the average energy of electronic neutrino is in the order of 1 MeV. It is possible to show that the rest-mass change of this neutrino type must be equal to approximately $\Delta m^{Th}_{\nu_e} \sim 4 \times 10^{-3}$ eV to account for loss of solar neutrinos due to oscillations. However, experiments for this neutrino type indicate $\Delta m_{\nu_e} \cong 4$ eV. An estimated electronic number density $n_e^{IS} \sim 5 \times 10^{36}$ m$^{-3}$ of the Sun's internal structures is necessary for observable deficit in the neutrino's rest mass. Neutrino dispersion is less probable, and we neglect this effect in our analysis. Consequently, resonant oscillations due to density gradients will be the basic mechanism for image formation of internal Sun structures. In other words, the Sun's neutrino flux already contains image information about the Sun's interior. Development for extraction of this information will be addressed in the following.

## IV. COPY OF NEUTRINO IMAGES IN ACCOMPANYING FLUX

An accompanying flux contains a duplicate of the neutrino image information as a result of interaction of the neutrino flux emanating from the Sun. In the following we discuss mechanisms that may lead to quanta created by the neutrino flux. As well as neutrons, possessing magnetic moment, neutrinos can generate quanta during interaction with a magnetic field. Neutrinos can deflect perpendicular to the direction of a magnetic field $\vec{H}$. At the radiation frequency $\omega$ the magnetic field [19] must be $H \sim \dfrac{\omega \hbar (1-\beta)}{2\mu_\nu^0}$. Here, $\hbar$ is Planck's constant divided by $2\pi$, and $\beta$ is the relativistic factor for solar electronic neutrinos. For example, electronic solar neutrinos with emitted quanta of wave-length $\lambda \sim 200$ Å imply a magnetic field of $H \sim 2 \times 10^{13}$ Tesla. In the Sun's environment such field strengths are not observed. Therefore, it becomes necessary to consider that these magnetic fields may exist only inside of filaments. These filaments should have direct and opposing magnetic flux. Moreover, the magnetic field does not exist outside of these filaments. The physical nature and modeling of these filaments will be discussed further below.

Another mechanism for generating accompanying/co-propagating quanta of the same frequency $\omega$ (corresponding to $\lambda \sim 200$ Å) is based on radiation from neutrinos. As neutrinos propagate they experience a spatially-varying magnetic field (caused by the filament sequence) along their trajectory with linear number density in the order of $n \sim 10^8$ m$^{-1}$. In this case accompanying quanta are radiated by a neutrino due to magnetic fields in the filaments.

Image formation inside the Sun and its analysis are connected with filament properties. The magnetic field can amount to $2 \times 10^{13}$ T inside the filaments. The electronic number density of the Sun can be as large as $5 \times 10^{36}$ m$^{-3}$ near internal-structure filaments. This implies that neutrino astronomy with high spatial resolution is based on filament properties.

## V. FILAMENT PARAMETERS AND STRUCTURES

The filament matter model [20] is still in its development stage but it is suitable for preliminary estimates of basic parameters. Filaments represent extended (infinite) chains of quarks in the Bose-condensate state connected to quantified magnetic field flux. The minimal flux value equals $\phi_0 = \pi h c / e$. Here $h$ denotes the Plank constant, $c$ the velocity of light in vacuum and $e$ the charge of electron. All quarks that are in the Bose-condensate state are relativistic and have identical energy of 5 MeV. The quark filament radius $r_q$ is $5 \times 10^{-11}$ m. This radius is obtained from the size of the deBroglie wave. The magnetic field strength (flux into one quant that connects the quarks of extended filament) equals typically $H_q \sim 3 \times 10^{13}$ T inside the filament. The energy of the quark connection is $\sim 5$ GeV.

The most important result of the presented analysis is the fractal nature of the discovered structure in the Universe. However, above described model of fractal matter does not include mechanisms that explain generation of these structures. Modifications of the model are based on assuming that fractal matter is formed from a very hot universe. The final-size quark filaments could show as well a shell of quarks that intercept the returning magnetic flux. Internal cords and external shells of extended filaments of such matter can develop consecutive chains of quarks with various combinations of charge composition. In particular, filaments can occur that consist of identical sequences of quarks in the Bose-condensate state, e.g., $4 \times (-1/3) + 2 \times (+2/3) + 4 \times (-1/3) + 2 \times (+2/3) + \cdots$. Filaments can exist also inside a star with a charged sequence of quarks, e.g., $2 \times (-1/3) + 2 \times (+2/3) + 2 \times (-1/3) + 2 \times (+2/3) + \cdots$. Outside shells one can find electrons in the Bose-condensate state as well. The minimal radius is $r_e \sim 3 \times 10^{-14}$ m for such a shell of relativistic electrons. The shell covers a returning magnetic flux of central cord. The field inside the shell is $H_e \sim 7 \times 10^{11}$ Tesla. The charge per unit length of a filament is equal to zero. Extended filaments of fractal matter can be assembled in separate cylindrical blocks connected by the fine structure constant $\alpha = 1/137$. The blocks of minimal length can viewed as "linear atoms" of fractal matter. One can estimate within the linear atom model the number of nucleons $N_N \sim (0.6 \text{ to } 1.2) \times 10^3$, the linear number density of nucleons $n_N \sim (0.6 \text{ to } 1.2) \times 10^{16}$ m$^{-1}$, its mass $m_{aq} \sim (1.2 \text{ to } 2.4) \times 10^{-24}$ kg, the length $l_a \sim 10^{-13}$ m, and a length-to-diameter ratio of $\sim 5$ with diameter of approximately $4 r_q \sim 2 \times 10^{-14}$ m. The fractal-matter density is $\sim (4 \text{ to } 8) \times 10^{16}$ kg/m$^3$, i.e., only approximately (10 to 5) times less than the density of typical nuclear matter. The linear atoms of fractal matter can form a fractal skeleton structure [1, 5]. With an average density of the sun $\rho_C \sim 10^3$ kg/m$^3$ the volume fraction of fractal matter is expected to be small. The linear-atom radius of fractal matter is $10^4$ times less than the radius of hydrogen atom. They are neutral and do not interact with usual atoms. Interaction of this matter with usual nuclei must also be weak. Fractal matter can exist in neutral state inside the Sun. Neutrality is extended to electrons in an external filament shell showing a connection-energy of the order of $\sim 5$ MeV. The linear-atom electron number density is $n_e \sim 3 \times 10^{15}$ m$^{-1}$, with mass $m_{ae} \sim 6 \times 10^{-25}$ kg, or approximately $m_{ae} \sim 3 \times 10^2$ m$_p$, where m$_p$ is the mass of a proton. The linear density amounts to $6 \times 10^{-12}$ kg/m, and the ratio of length and of diameter is typically 2.

The discovered skeleton structure of the Sun shows a tendency towards self-similarity, i.e., towards a fractal structure. The topology was found identical to the one previously observed in dusty carbon deposits that were extracted from plasma chambers. The model for the fractal skeletal structures in fractal matter of linear atoms is similar in construction to the one that describes carbon-containing nano-tubes [7-9, 23]. The mass, number of electrons, length and diameter for such $n$-th generation tubes is completely determined. Diameter and length can be estimated from: $d_n \sim 6 \times 10^{-14} \times 5^{n-1}$ m and $l_n \sim 2d_n \sim 12 \times 10^{-14} \times 5^{n-1}$ m. Blocks of various growth size of the fractal skeleton structure show a scale factor K = 5. The 2-nd generations tubular filaments show radial connections of typically 80 linear atoms. The number of linear atoms is $N_L^n \sim (80)^{n-1} = 2^{3(n-1)} \times 10^{n-1}$ for a $n$-th generation filament. The total number of electrons $N_{Le}^n \sim N_N \cdot N_L^n = 3 \cdot 10^2 \cdot 2^{3(n-1)} \cdot 10^{n+1}$ is contained in the filament with volume $V_L^n \sim 0.25 \times \pi \times d_n^2 \sim 4.3 \times 10^{-40} \times 5^{3(n-1)}$ m$^3$. The average electron number density of the filament amounts to $n_{Le}^n \sim N_{Le}^n / V_{Le}^n \approx 2^{4n+32}/5^{2n-38} \times 10^6$ m$^{-3}$. Filaments of fractal matter show the characteristic diameter of $3 \times 10^{-11}$ m inferred from images of the Sun [18]. Consequently, we find a value of $n \sim 30$ for the generation number. Using above formula we compute $n_{Le}^n \sim 2.4 \times 10^{36}$ m$^{-3}$ for the average electron number density of the filaments. This value is in agreement with the one obtained in the discussion of neutrino oscillations (see Section III).

It is not necessary to average the magnetic field over the volume of a layer occupied by the fractal matter because a generation with $n = 30$ shows a diameter slightly larger than the Sun diameter. However, it is difficult to obtain a dense enough flux of accompanying/co-propagating quanta for visualization of the inside of the Sun. (As discussed before, the accompanying/co-propagating quanta are generated due to interaction of solar flux neutrinos with magnetic field of the fractal matter located near but outside the surface of the Sun.) Our estimates confirm experimental difficulties when using the presented model for skeleton-structure growth when considering collection of data during several tens of minutes with a telescope entrance surface of ~ 1 m$^2$.

Accompanying/co-propagating quanta may be radiated if neutrinos propagate in strong and variable magnetic fields. The linear atoms of the fractal matter carry direct and returning magnetic, quantized flux. Neutrino motion is across the fractal matter. For a sufficiently strong, varying magnetic field the neutrino motion across the fractal matter can be calculated from the average of the linear atoms of the fractal matter along the trajectory. The oriented linear-atom layer of the fractal matter can be comprised of free linear atoms. These free linear atoms are omitted in the rotating fractal-skeleton-structure of the Sun but are responsible for generation of a halo at a distance from the sun determined using equality of gravitation and centrifugal force. We estimate that the halo will be located at a distance of approximately ~ $10^{11}$ m from the Sun. Images from the Sun discussed in this work are obtained at a wavelength of typically $10 \times 10^{-9}$ m; therefore, the average linear number density of the linear atoms in fractal matter that can be inferred should be of the order of $0.1 \times 10^9$ m$^{-1}$. It would be reasonable to include neutrino interaction with probable "dark matter" halo in considerations of possible mechanisms for latent images in neutrino flux measurements. Inclusion of dark-matter effects in halo estimates would require knowledge of the physical nature of dark matter. However, dark matter interactions continue to remain a puzzle for us, equally the interactions with fractal-skeleton-structures and with neutrinos.

It is noteworthy that the linear atoms of the fractal matter can be also considered as dark-matter particles and, similarly, fractal skeleton structures can exist in dark matter as well. Should confirmation occur of the presented inferences about fractal-skeleton-structure, then we could also infer validity of the conversion and image-construction mechanisms from a neutrino flux that contains information about the Sun's interior. Nevertheless, it will be necessary to explore details of the mechanism for generating accompanying/co-propagating quanta during neutrino propagation. Current estimates show that the flux of accompanying x-ray quanta should be sufficient for generation of images of the Sun's interior. The estimates were obtained from an analysis of a database of images of the Sun, and by considering collection of quanta by a telescope with an entrance lens size of typically 1 $m^2$ for tens of minutes of exposure.

At the same time the spatial resolution can reach values up to $5 \times 10^5$ m that is almost $10^4$ times higher than can be obtained with the state-of-the-art detector, named Super-KamiokaNDE for Kamioka Nucleon Decay Experiment (NDE) [24]. Detailed consideration of the given problem is complex because all available calculations use radiation probability of quanta in a homogeneous magnetic field. As neutrinos pass across fibers of fractal matter high magnetic field strengths of fractal matter can induce a significant radiation probability, thereby increasing the number of accompanying/co-propagating quanta that can be measured.

## VI. SUPER-KAMIOKANDE NEUTRINO TELESCOPE

A neutrino telescope has been designed on the principle of directly detecting neutrinos. Recent experiments utilize Japan's neutrino telescope with the Super-KamiokaNDE detector [24]. The measurement mechanism of this telescope is based on transfer of energy from energetic neutrinos to electrons along the trajectory of the neutrinos in the detector. The electrons emit Cerenkov radiation into a cone of light in the direction of their propagation. Cerenkov radiation is registered by sets of spatially-distributed, sensitive photomultipliers. A cylindrical tank filled with pure water is at the base of a telescope. Its height is 36 m with a diameter of 34 m. The full weight of a tank is $50 \times 10^9$ kg. Walls and the bottom of the telescope tank hold an assembly of 11146 specially-designed photomultipliers with low-noise photo-cathodes of 0.5-m diameter. The photomultipliers are arranged such that the entire volume can be readily monitored. An image of the Sun has been recorded with an almost 3-year exposure time after the telescope was built. Figure 7 illustrates results collected with the neutrino telescope. The neutrino profile of the Sun was obtained using the Super-KamiokaNDE detector [24] and enhanced with the multilevel dynamic contrast (MDC) method. One pixel of the image corresponds to one degree which is to be compared with the one-half degree for the solar disk. Contrast distribution can be determined from neutrino dispersion.

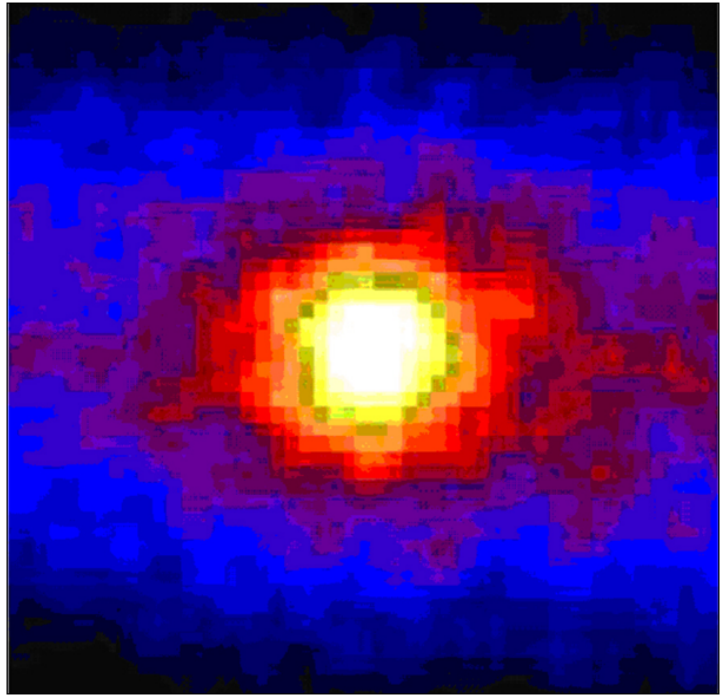

**Figure 7.** Neutrino image of the Sun.

### VI.     NEUTRINO ASTRONOMY WITH THE HIGH SPATIAL RESOLUTION

Solar Dirac-neutrinos possess non-zero magnetic moment and radiate quanta of light during interaction with a magnetic field. Typically large magnetic field strengths are required for this process to occur because the magnetic moment of the neutrino is relatively small. Field strengths inside filaments are strong enough for deflection of Dirac-neutrinos. Neutrino astronomy at high spatial resolution would be possible if the suggested fractal matter is indeed confirmed. Interior images of stars can be generated using a neutrino flux that originates from the same star. Furthermore, this neutrino flux can carry information for image formation to the outside of the star because neutrinos do interact (i) with the fractal-skeleton-structure of fractal matter of the Sun and (ii) with the halo of oriented, free linear-atoms of dark matter. Decoding of the image would occur by the detector of the telescope using accompanying/co-propagating quanta generated from neutrinos. Presence of a halo can explain occurrence of a horizontal strip in the neutrino-flux image of a solar disk (see Fig. 7). Neutrinos scatter from the region of the indicated strip. From geometrical optics one can infer that the vertical extent of the strip corresponds to the distance of the halo from the observer.

Optical or x-ray images show high spatial resolution. Neutrino astronomy discussed above can occur provided fractal matter exists as described in Section V. Examples presented in this paper show that the suggested hypotheses can indeed be substantiated, namely existence of fractal-skeleton-matter inside and in the vicinity of the Sun.

Images that indicate the internal structure of the Sun can help resolve problems in neutrino physics: a) Existence of neutrino oscillations, i.e., solar neutrinos are Dirac-neutrinos; b) Observation of fractal-skeleton-matter including dynamics inside stars/galaxies from analysis of images recorded at various wave lengths; c) Dynamic character of the neutrino moment and mass including interactions with linear-atoms of fractal and/or dark matter particles.

High spatial resolution neutrino astronomy with existing space telescopes is now possible. It allows us to look into centers of stars and galaxies, to observe their internal structure and to study processes taking place in the interior, to reveal fractal skeleton matter and to study their properties. Currently existing space telescopes show an enormous database of data collected over several decades. This database appears to already contain information from the interior of stars and galaxies. However, it is necessary to introduce new methodologies in interpreting the recorded images in view of how these images were formed and how information can be extracted. For example, a flash from a star of the nearest galaxy to us, as was shown recently, should not necessarily be connected with "gravitational lensing" caused by a black hole. The observed phenomena can be possibly explained by an additional flux of quanta caused by a star's neutrino flux that propagates in space between star and observer and that travels across filaments of fractal matter. Fractal matter discoveries can motivate new initiatives for space research, can help understand formation processes of cosmic objects, and can assist in the search for new energy sources in the universe.

## VII.   STRUCTURES OF LUMINOUS OBJECTS IN THE UNIVERSE

Available maps of red-shift of galaxies and quasars have also been explored in view of potentially embedded fractal structure. This work indeed extends the fractal picture to larger spatial scales in our Universe. The multilevel-dynamical-contrast method was again applied in this effort. Analysis of images (obtained for various wave length ranges) from luminous objects in the universe allows us to infer novel hypotheses.  One of the new phenomena which were discovered from images of laboratory and cosmic plasma, were rectilinear dark filament structures with ends that can shine like ends of optical fibers. The wavelength region of the light emission corresponds to the temperature of the studied plasma. These coaxial-tubular structures were named "electric torch-like structures" [5, 25]. Figure 8 illustrates images from the tokamak T-6. Fig. 8a) illustrates electric torch-like structures of tokamak T-6 luminescence for a 1 μs exposure time. The image height corresponds to 0.08 m. The lower left corner shows CTS of 0.01 m diameter that extends almost diagonally. Fig. 8 b) elucidates hotter and denser plasma. A vertically aligned plasma column occurs in the image (width 1.75 cm) of the electrical discharge of the Z-pinch type. A dashed-dotted line shows the Z-pinch axis. The image (positive) is recorded in visible light with a 2 nano-second exposure time while the discharge's magnetic field squeezes the hot plasma column and partially strips a skeletal network from the ambient luminous plasma (representing  an example of a strongly stripped skeleton in the same Z-pinch).  The 3.5 milli-meter wide window indicates that the "hot spot" is at the filament's edge which is close to a brighter core as well.

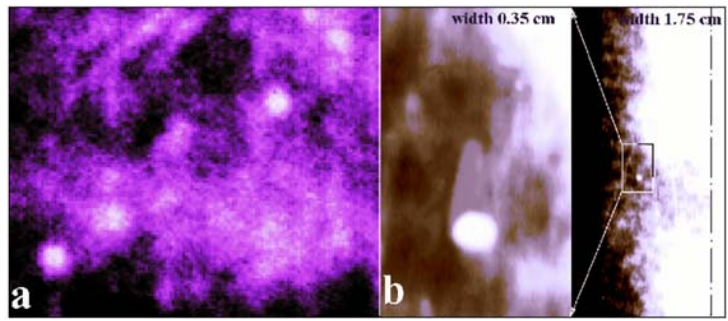

**Figure 8.** Tokamak T-6 images. **a.** Torch-like structures collected during exposure time of 1 μs. **b.** Image of denser and hotter core of Z-pinch type discharge.

It appears that straight blocks of the skeleton may behave as guiding systems for conduction of electromagnetic signals. Therefore, the open end of a dendrite electric circuit or a local disruption of such a circuit (viz. sparks, fractures, etc.) may cause self-illumination. In turn, this self-illumination allows us to observe these structures. Similar phenomena can occur in space plasma. *Many luminous objects in the Universe appear as luminous ends of the coaxial-tubular structures*. The spatial extend of the ends of such open optical paths can be associated with the sizes of stars, planetary nebulas, or galaxies. Figures 9 and 10 show examples of electric torch-like structures observed in space. Fig. 9 shows torch-like structure of the 0.1 light-year diameter filament in a section of the optical image of the Crab nebula [26]. The light from the filament's edge anisotropically illuminates the ambient gas. In spite of the cylindrical body the filament is visible due to either an internal radiation source or external illumination. The visibility of the filament is not diminished by the partially opaque surrounding medium indicated in the left upper corner of the window.

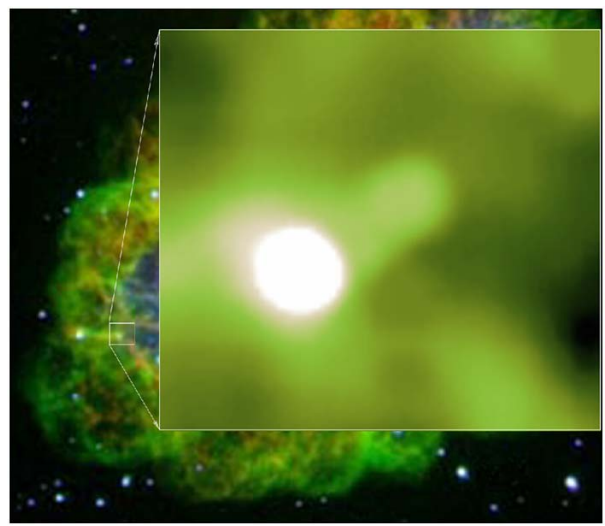

**Figure 9.** Optical Image of the Crab nebula.

Fig. 10 illustrates the electric torch-like structure in the 2,400 light-year-wide core of the spiral galaxy NGC 1512 [27] which is 30 million light-years away from us and shows a 70,000 light-years lateral extent. A color-composite image and one obtained with 2200 Å optical radiation are displayed together with the enhanced image obtained with help of the multilevel dynamical contrast method.

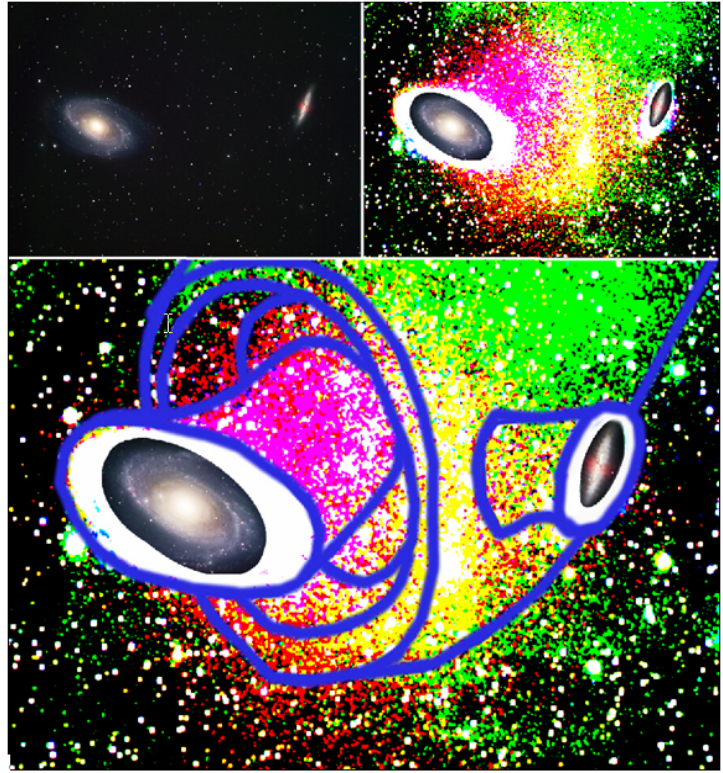

**Figure 10.** Images of the spiral galaxy NGC 1512. **a.** Color composite image. **b.** Image obtain at 220-nm. **c.** Enhanced image.

Image analysis of many space objects leads to the conclusion that the majority of observable objects in the universe are luminous ends of almost invisible (or almost completely transparent) formations of filaments. The structure of these filaments can be seen and/or inferred only near their luminous ends. We only "see" the luminous ends which in turn can be the reason that we see in the Universe what amounts to be only a few percents of the entire mass. One can show not only luminous objects of galaxies, but also galaxies themselves are luminous open ends of complex structure of fractal matter. This fractal matter is practically "dark matter" because it manifests itself only in gravitational interaction. The large-scale coaxial-tubular structure of the fractal matter, or dark matter, can be interconnected in a unique network of the Universe. All objects of this network are directly connected. Neighboring galaxies are especially strongly connected when they belong to a single tree-like filament, and they appear as ends of cut-off branches. Occasionally one can recognize this scenario. For example, interacting galaxies (M 81 and M 82) appear as ends of two cut-off branches of a treelike filament; their interaction protrudes through this filament. Figure 11 shows these phenomena for the M81 and M82 galaxies, members of the "The Big she-bear" constellation [28]. Fig. 11b) also shows image-enhanced results. Application of the multilevel dynamical contrast method allowed us to discover details of their interaction. M81 appears as a cut-off trunk of a tree-like filament of $3.5 \times 10^{21}$ m diameter, and M82 appears as the end of a cut-off lateral branch. The figure shows M82 on the right with the basic filament of dark CTS that resembles a cut-off branch. The newly created specific galaxy is shown parallel to the axis. The bright luminous object located slightly below but in front of M82 is located on an axis of this dark structure. Noteworthy is the following: *Contemporary astronomers may explain the motion of each galaxy as an indication of cosmic expansion. However, contrary to this interpretation the image shows that the galaxies indeed collided.*

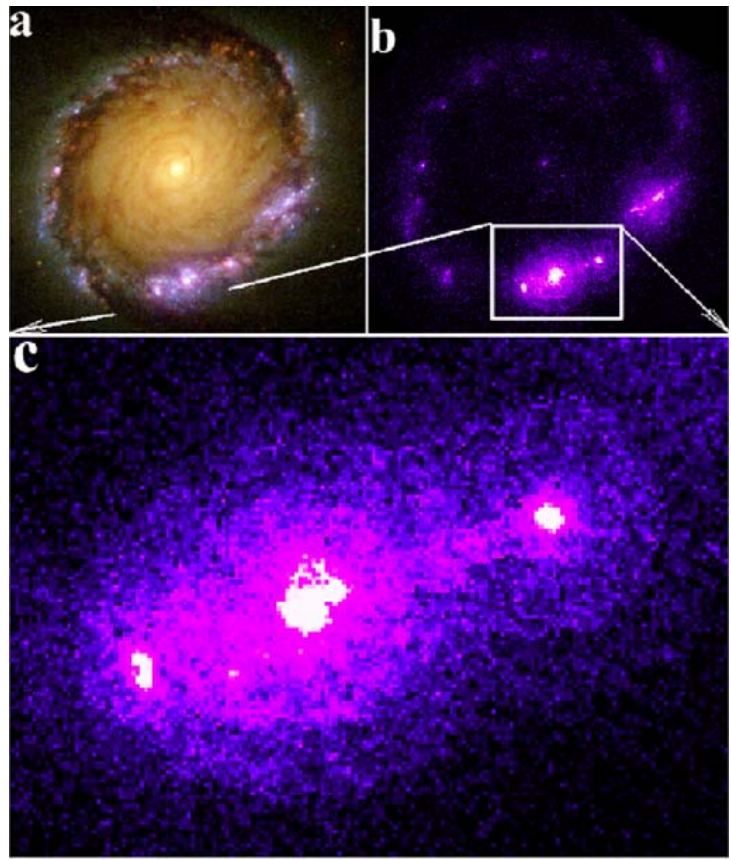

**Figure 11.** Intersecting M81 and M82 galaxies. **a.** Original image. **b.** Enhanced image.

Similar multilevel-dynamical-contrast analyses of images from Earth, Venus, and Saturn, taken in the visible light spectrum by Voyager 1 outside the solar system, show that these planets are located on dusty filaments directed perpendicular to the ecliptic plane (see Figs. 8–10 in Ref. [29]). The coaxial-tubular filaments are found to be of nearly identical size and are directed almost perpendicular to the axis of the filament. The planets are located at the ends of the tubular blocks in the ecliptic plane. The images show that among the blocks composed of filaments, only those are in the ecliptic plane that possess a planet at its ends. This observation suggests that the planets of the solar system are contained on a filament that developed from cosmic dust due to gravitational interaction. One can infer that the solar system was formed on an end of a ruptured filament. This particular filament would be of the dark-matter kind with diameter corresponding to the size of the solar system.

### VIII. BIRTH OF GALAXIES AND UNIVERSE

Cosmic objects show frequently a structure of paired interacting galaxies. The galaxies appear as two opposing ends reminiscent of a filament break. The two separated fragments are mutually interconnected near and/or at the breakage region. When applying the multilevel-dynamic-contrast (MDC) algorithm, silhouettes of filaments become clearly recognizable in the images. Our Universe shows a steady dynamical state of creation of stars and/or galaxies. Their density can increase resulting from rupture of a filament with a spatial extent that corresponds to the diameter of the cosmic objects. Formation of galaxies occurs by means of the mentioned filament-breaking mechanism. The time scale for this "breaking" process is much smaller that the time-sale predicted from standard models that are based on consideration of gravitation instabilities.

The "wheels" are most representative in the common structure of the Universe. Their dynamics leads to observation of the role of these structures in cosmic events. It is possible to show, *all interacting galaxies represent breaks and/or collisions of similar wheel structures (of corresponding size).* Figure 12 shows the recorded and contrast-enhanced image from two interacting galaxies, UGC 06471 and UGC 06472.

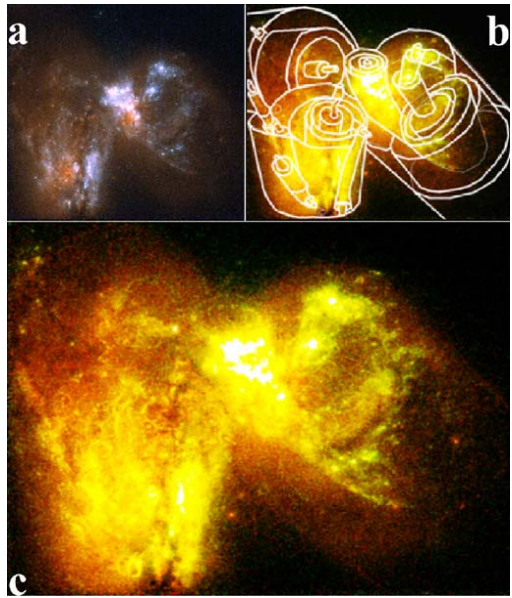

**Figure 12. a.** Images of interacting galaxies UGC 06471 and UGC 06472. **b.** Image-enhanced schematic. **c.** Enhanced image at a magnified scale.

Fig. 12 also shows an illustration of the tubular structures. The original image for the interacting galaxies UGC 06471 and UGC 06472 was recorded in the uv [30]. Both **s**chematic representation of the image, and the MDC enhanced image illustrate interaction of three CTSs. It can be inferred that two galaxies in the foreground are ends of a rupture of one of the coaxial-tubular structures. A third CTS appears to have collided with a joint of the first two, in turn, allowing us to infer that the third CTS may be the reason for the rupture to occur. Detailed analysis of this CTS group shows their topological identity. These blocks show telescopically enclosed tubes with radial connections.

Fig. 13 shows images of two interacting galaxies NGC 4038/4039 from the constellation Corvus [31]. The figure also shows the enhanced image together with detailed cartoons of the separated galaxies. These two galaxies show the fractal topology. The structure of the top galaxy shows radial spokes which connect the central coaxial-tubular part of the galaxy with tubes located on lateral surface of a filament. Details of the lower and upper galaxy illustrate the same topology as for the entire galaxy. Most importantly, the indicated connections can be matched precisely! This would imply that birth is caused by breakage of a single structure, namely a coaxial-tubular structure.

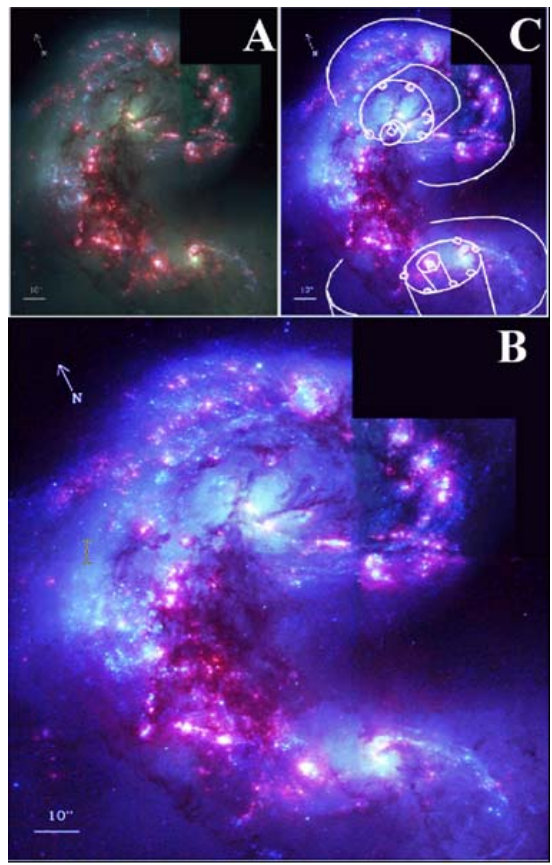

**Figure 13. a.** Two interacting galaxies NGC 4038/4039. **b.** Enhanced images.

It is reasonable to ponder the following: if we were capable to analyze expanding structures at spatial scales larger than size of our Universe then we perhaps would see the same topology in a so-called mega-universe comprised of Universes similar to our Universe. Consequently, in analogy to birth of galaxies discussed earlier, one can infer that Universes are created following ruptures of dark-matter filaments of spatial extend of the order of our Universe.

### IX. CONCLUSIONS

The main conclusions are summarized here. Firstly, the Universe shows the universal topology skeletal structure which consists of separate coaxial-tubular structures and wheel-like building blocks. Secondly, the Universe shows self-similarity at any scale, and hence is of fractal nature. Thirdly, luminous objects observable in the Universe can be associated with free ends or with filament ruptures of corresponding size of coaxial-tubular structures. Fourthly, the Universe continues to be in a dynamic state of forming stars and/or galaxies. The current distribution can be explained by means of fracturing filaments. The formation time of cosmic objects is shorter than predicted by the standard model. The structure of the objects is determined by the filament structure itself. The formation time of a given filament is equal to the one for a baryonic mass in multiples of three quarks of the given filament. The new cosmic objects are ejected from the filament breaking point in form of nucleons.

Several examples discussed in this work indicate that the Universe shows the same topology over a wide range of scales. The topology is now observed over spatial scales from $10^{-9}$ m up to $10^{26}$ m, i.e., over 35 orders of magnitude. One may expect that mega-universes show similar topology if we only would have the capability to analyze structures bigger than our Universe. According to the mentioned mechanism for formation/birth of galaxies, it is entirely plausible to consider that Universes may be created due to rupture of dark-matter filaments of comparable size. This includes the inference that our Universe was created by dark-matter filament rupture.

**Acknowledgments**

One of us (VAR-K) thanks A.B. Kukushkin for his decade-long collaboration. Special thanks go to V.I. Kogan for invariable support and interest in research of skeletal fractal structures. CGP acknowledges support in part from the Russian Academy of Natural Sciences and in part from the Center for Laser Application to attend the 2010 Plasma Science conference in Zvenigorod, Russia, that allowed us to commence international collaboration on this subject.